\documentstyle{article}
\begin{document}
\begin{titlepage}
\centerline{\LARGE Fixed Point of the Finite System DMRG}
\vskip 40pt

\centerline{ Hiroshi {\sc Takasaki},
Toshiya {\sc Hikihara},$^{1}$ and Tomotoshi {\sc Nishino}$^{2}$}
\centerline{\sl Department of Physics, Graduate School of Science and
Technology,}
\centerline{Kobe University, Rokkodai 657-8501}
\centerline{\sl $^1$Division of Information and Media Science,
Graduate School of Science and
Technology,}
\centerline{Kobe University, Rokkodai 657-8501 }
\centerline{\sl $^2$Department of Physics, Faculty of Science, 
Kobe University, Rokkodai 657-8501}
\vskip 40pt

\begin{abstract}
The density matrix renormalization group (DMRG)
is a numerical method that
optimizes a variational state expressed by a tensor product. We show that
the ground state is not fully optimized as far as we use the standard finite
system algorithm, that uses the block structure 
${\bf B} \bullet \bullet  {\bf B}$.
This is because the tensors are not improved directly. We overcome
this problem by using the simpler block structure ${\bf B} \bullet  {\bf B}$
for the final several sweeps in the finite iteration process. It is possible to
increase the numerical precision of the finite system algorithm 
without increasing the computational effort. 
\end{abstract}
\end{titlepage}

Establishment of the density matrix renormalization group (DMRG) by
White~\cite{Wh1} is one of the major progresses in computational
condensed matter physics. DMRG enables us to calculate ground
states of relatively large scale one-dimensional (1D) quantum
systems.~\cite{WF,SA,corr,Geh,Escorial}. Two-dimensional (2D) classical
systems,~\cite{Ni,Carlon1,Carlon2,Carlon3} and 1D quantum system at
finite temperature~\cite{Bur,Wang,Shibata,Egg} have also been 
investigated.

\"Ostlund and Rommer~\cite{Ost} examined the thermodynamic limit ($N
\rightarrow \infty$) of the infinite system algorithm, and they
pointed out that the block state ${\bf B}$ corresponds to a product of
position independent tensor. It should be noted that
their result does not show that the infinite system algorithm creates a
translationally invariant --- position independent --- variational state for
the whole system ${\bf B} \bullet \bullet  {\bf B}$, 
where ``$\bullet$'' denotes a bare
spin variable between the left and the right blocks. Actually,
the variational state has a slight position dependence.
For example, the bond energy $\langle \bullet \bullet  \rangle$
at the center of the antiferromagnetic $S = 1/2$ Heisenberg chain,
which is calculated by the infinite system algorithm, is lower than the exact
ground state energy per site.~\cite{bnd} Such a position dependence in the
variational state spoils the numerical efficiency of the infinite system
algorithm.~\cite{bndprb} As we show in the following, the finite system
algorithm does not fully improve the variational state in the same reason.
The purpose of this letter is to remove the source of such a numerical
error, and to increase the numerical precision in DMRG. 

Let us briefly review the construction of the variational state, which is
used in the standard
finite system algorithm. We consider the IRF model~\cite{Bx1}
as a reference
system, whose transfer matrix is written as the product of local
Boltzmann weights
\begin{equation}
T_{s_{1\cdots}^{~}s_N}^{s'_{1\cdots}s'_N} \, = \, 
W_{s_1^{~}s_2^{~}}^{s'_1 s'_2} \,\,
W_{s_2^{~}s_3^{~}}^{s'_2 s'_3} \cdots
W_{s_{N-1}^{~}s_N^{~}}^{s'_{N-1}s'_N} \, ,
\end{equation}
where $W$ represents the IRF weight. The finite system
algorithm, that uses the block structure ${\bf B} \bullet \bullet  {\bf B}$,
approximates the eigenvector of 
the transfer matrix using a variational state in the tensor product 
%
% ********** Fig. 1a ************
% 
%
\begin{eqnarray}
V_{s_{1\cdots}^{~}s_N^{~}}^{(M)} = 
\!\!\!\!\! \sum_{\xi_2^{~}\cdots\xi_{N-1}^{~}}^{{\rm at~most~}m} \!\!\!\! 
&&
A_{s_1^{~}\xi_2^{~}}^{s_2^{~}} \cdots
A_{\xi_{M-2}^{~}\xi_{M-1}^{~}}^{s_{M-1}^{~}}
{\tilde V}_{\xi_{M-1}^{~}\xi_{M+2}^{~}}^{s_{M}^{~}~~s_{M+1}^{~}} 
\nonumber\\
&&
B_{\xi_{M+2}^{~}\xi_{M+3}^{~}}^{~~~~~~s_{M+2}^{~}} \cdots
B_{\xi_{N-1}^{~}s_{N}^{~}}^{~~~~~~s_{N-1}^{~}} \, , 
\end{eqnarray}
where $\xi_2^{~}\cdots\xi_{N-1}^{~}$ denote $m$-state block spin
variables. (Fig.1a)
The tensors $A_{\xi_{i-1}^{~}\xi_i^{~}}^{s_i^{~}}$ and 
$B_{\xi_j^{~}\xi_{j+1}^{~}}^{~~~s_j^{~}}$ are dependent on their positions $i$
and $j$, and each of them satisfies the orthogonal relation
\begin{eqnarray}
\sum_{\xi_{i-1}^{~}s_i^{~}}^{~}
A_{\xi_{i-1}^{~}\xi_i^{~}}^{s_i^{~}}
A_{\xi_{i-1}^{~}\xi'_i}^{s_i^{~}} 
\, &=& \, \delta_{\xi_i^{~}\xi'_i}^{~} \nonumber\\
\sum_{s_j^{~}\xi_{j+1}^{~}}^{~}
B_{\xi_j^{~}\xi_{j+1}^{~}}^{~~~s_j^{~}}
B_{\xi'_j\xi_{j+1}^{~}}^{~~~s_j^{~}}
\, &=& \, \delta_{\xi_j^{~}\xi'_j}^{~} \, ,
\end{eqnarray}
where we have written $s_1^{~}$ and $s_N^{~}$ as
$\xi_1^{~}$ and $\xi_N^{~}$, respectively.
Normally, they impose the normalization~\cite{sym}
\begin{equation}
\sum_{\xi_{M-1}^{~}s_{M}^{~}s_{M+1}^{~}\xi_{M+2}^{~}}^{~}
\!\!\!\!
\left( {\tilde V}_{\xi_{M-1}^{~}\xi_{M+2}^{~}}^{s_{M}^{~}~~s_{M+1}^{~}}
\right)^2_{~} \, = \, 1 \, .
\end{equation}

The standard finite system algorithm improves the variational state
(Eq.2) so that (Fig.1b)
\begin{equation}
\lambda_{~}^{(M)} = \sum_{\rm all~indices}^{~} 
V_{s'_{1\cdots}s'_N}^{(M)} \,
T_{s_{1\cdots}^{~}s_N}^{s'_{1\cdots}s'_N} \,\,
V_{s_{1\cdots}^{~}s_N^{~}}^{(M)}
\end{equation}
is maximized via the tuning of ${\tilde
V}_{\xi_{M-1}^{~}\xi_{M+2}^{~}}^{s_{M}^{~}~~s_{M+1}^{~}}$, under the
constraint Eq.4. 
%
% *********** Fig.1b ***********
%
The best ${\tilde V}_{\xi_{M-1}^{~}\xi_{M+2}^{~}}^{s_{M}^{~}~~s_{M+1}^{~}}$ 
is determined by the diagonalization of the renormalized transfer matrix
\begin{equation}
{\tilde T}_{\xi_{M-1}^{~}s_M^{~}s_{M+1}^{~}\xi_{M+2}^{~}}^{
\xi'_{M-1}s'_M s'_{M+1}\xi'_{M+2}}
=
{\tilde L}_{\xi_{M-1}^{~}s_M^{~}}^{\xi'_{M-1}s'_M}
W_{s_M^{~}s_{M+1}^{~}}^{s'_M s'_{M+1}}
{\tilde R}_{s_{M+1}^{~}\xi_{M+2}^{~}}^{s'_{M+1}\xi'_{M+2}} \, ,
\end{equation}
and by identifying its eigenvector 
${\tilde V}_{\xi_{M-1}^{~}s_M^{~}s_{M+1}^{~}\xi_{M+2}^{~}}$
with the tensor
${\tilde V}_{\xi_{M-1}^{~}\xi_{M+2}^{~}}^{s_{M}^{~}~~s_{M+1}^{~}}$.
The factor
${\tilde L}_{\xi_{M-1}^{~}s_M^{~}}^{\xi'_{M-1}s'_M}$ and
${\tilde R}_{s_{M+1}^{~}\xi_{M+2}^{~}}^{s'_{M+1}\xi'_{M+2}}$
represent the renormalized half-row transfer matrix
for the left and the right half of the system, respectively. (Fig.1c) 
%
% ********* Fig.1c ***********
%
%
A pair of tensors 
$A_{\xi_{M-1}^{~}\xi_M^{~}}^{s_M^{~}}$ and 
$B_{\xi_{M+1}^{~}\xi_{M+2}^{~}}^{~~~~~~s_{M+1}^{~}}$
are then improved {\it indirectly} by rewriting the
improved tensor ${\tilde
V}_{\xi_{M-1}^{~}\xi_{M+2}^{~}}^{s_{M}^{~}~~s_{M+1}^{~}}$
using the singular value decomposition (SVD)
\begin{equation}
{\tilde V}_{\xi_{M-1}^{~}\xi_{M+2}^{~}}^{s_{M}^{~}~~s_{M+1}^{~}}
= \sum_{\xi_M^{~}\xi_{M+1}^{~}}^{2m}
A_{\xi_{M-1}^{~}\xi_M^{~}}^{s_M^{~}} \, 
\Omega_{\xi_M^{~}\xi_{M+1}^{~}}^{~} \, 
B_{\xi_{M+1}^{~}\xi_{M+2}^{~}}^{~~~~~~s_{M+1}^{~}} \, ,
\end{equation}
and by restricting the degree of freedom of
$\xi_M^{~}$ and $\xi_{M+1}^{~}$ down to $m$. The matrix
$\Omega_{\xi_M^{~}\xi_{M+1}^{~}}^{~}$ is a $2m$-dimensional
diagonal matrix~\cite{notd}
\begin{equation}
\Omega = \left(
\begin{array}{cccc}
\omega_1^{~} & ~ & ~ & ~ \\
~ & \omega_2^{~} & ~ & ~ \\
~ & ~ & \ddots & ~ \\
~ & ~ & ~ & \omega_{2m}^{~}
\end{array}
\right) \, ,
\end{equation}
where the diagonal elements are in the decreasing order
$|\omega_1^{~}| \geq |\omega_2^{~}| \geq \cdots
|\omega_{2m}^{~}|$.  The matrix $\Omega$ satisfies the normalization
${\rm Tr} \, \Omega^2_{~} \, = \, \sum_{\xi}^{2m} \omega_{\xi}^2 \, 
= \, 1$. The finite system algorithm improves other tensors in Eq.2
by shifting the position of ${\tilde V}$ by use of the wave function
renormalization.~\cite{Whacce,pwfrg}

In the above standard improvement process for 
the variational state
$V_{s_{1\cdots}^{~}s_N^{~}}^{(M)}$, the
tensors $A_{\xi_{M-1}^{~}\xi_M^{~}}^{s_M^{~}}$ and 
$B_{\xi_{M+1}^{~}\xi_{M+2}^{~}}^{~~~~~~s_{M+1}^{~}}$ are improved
indirectly, only through the tuning for ${\tilde
V}_{\xi_{M-1}^{~}\xi_{M+2}^{~}}^{s_{M}^{~}~~s_{M+1}^{~}}$. As a result,
$A_{\xi_{M-1}^{~}\xi_M^{~}}^{s_M^{~}}$ and 
$B_{\xi_{M+1}^{~}\xi_{M+2}^{~}}^{~~~~~~s_{M+1}^{~}}$
are determined under the condition where $2m$ degrees of 
freedom is allowed for both $\xi_M^{~}$ and $\xi_{M+1}^{~}$,
although only $m$ states are allowed for other block spin variables
$\xi_2^{~}\cdots\xi_{M-1}^{~}$ and $\xi_{M+2}^{~}\cdots\xi_N^{~}$.  It
is apparent that  additional $m$ numbers of freedom is allowed at the
position where ${\tilde V}$ is. 
(This excess freedom
is common to both the finite and the infinite system algorithms.) 
Thus the variational state $V_{s_{1\cdots}^{~}s_N^{~}}^{(M)}$
is dependent on $M$, even after many sweeps of the finite system
process. Figure 2 shows ${\rm ln} \left(\lambda_{~}^{(M)}\right)$ 
of the square lattice Ising model of the 
width $N = 200$ with free boundary condition, 
where we define the local Boltzmann weight as
\begin{eqnarray}
\lefteqn{W^{s'_is'_{i+1}}_{s_i^{~}s_{i+1}^{~}} }\\
&&= \exp \left\{ \frac{K}{2}
( s'_is'_{i+1} + s_i^{~}s_{i+1}^{~} + s_i^{~}s'_i + s'_{i+1}s_{i+1}^{~} - 4) 
\right\} \nonumber
\end{eqnarray}
for the Ising spins $s = \pm 1$.
We keep $(m =) 8$ states for the block spins.~\cite{kept}
We have chosen the critical temperature
$K = J / k_{\rm B} T_{\rm c}^{~}$.
Since $2m$ degrees of freedom is allowed 
for both $\xi_M^{~}$ and $\xi_{M+1}^{~}$,
the eigenvalue $\lambda_{~}^{(M)}$ of the renormalized transfer matrix
${\tilde T}_{\xi_{M-1}^{~}s_M^{~}s_{M+1}^{~}\xi_{M+2}^{~}}^{
\xi'_{M-1}s'_M s'_{M+1}\xi'_{M+2}}$ is dependent on $M$;
$\lambda_{~}^{(M)}$ takes its maximum when $M = N/2$.

The $M$ dependence of the variational state 
$V_{s_{1\cdots}^{~}s_N^{~}}^{(M)}$ causes an ambiguity for
the observation of local quantities. For example, 
they calculate the local magnetization
of the Ising model using the formulation
\begin{eqnarray}
\langle s_M^{~} \rangle 
&=& 
\sum_{s_{1\cdots}^{~}s_N^{~}}^{~}
V_{s_{1\cdots}^{~}s_N^{~}}^{(M)} \, s_M^{~} \, 
V_{s_{1\cdots}^{~}s_N^{~}}^{(M)} \\
&=& 
\!\!\!
\sum_{\xi_{M-1}^{~}s_{M}^{~}s_{M+1}^{~}\xi_{M+2}^{~}}^{~}
\!\!\!
{\tilde V}_{\xi_{M-1}^{~}\xi_{M+2}^{~}}^{s_{M}^{~}~~s_{M+1}^{~}}
\,\, s_M^{~} \,\,
{\tilde V}_{\xi_{M-1}^{~}\xi_{M+2}^{~}}^{s_{M}^{~}~~s_{M+1}^{~}} 
\nonumber \, ,
\end{eqnarray}
and therefore $\langle s_M^{~} \rangle$ and
$\langle s_{M'_{~}}^{~} \rangle$ for $M \ne M'_{~}$ are
calculated for {\it different} variational states
$V_{s_{1\cdots}^{~}s_N^{~}}^{(M)}$ and
$V_{s_{1\cdots}^{~}s_N^{~}}^{(M')}$, respectively.
The way to avoid such an ambiguity is simply to
obtain a variational state that is independent on $M$.

Now we show that we can further improve the variational state
using the block structure ${\bf B} \bullet {\bf B}$, and that the
improved variational state is not dependent on $M$.
The block structure ${\bf B} \bullet {\bf B}$
is known from the establishment
of DMRG,~\cite{Wh1,Xiam} but the difference between
${\bf B} \bullet \bullet   {\bf B}$ and ${\bf B} \bullet {\bf B}$ 
has not been investigated from the view point of the
numerical precision.
In this case, the renormalized transfer matrix is constructed as
\begin{equation}
{\tilde T}_{\xi_{M-1}^{~}s_M^{~}\xi_{M+1}^{~}}^{
\xi'_{M-1}s'_M \xi'_{M+1}}
=
{\tilde L}_{\xi_{M-1}^{~}s_M^{~}}^{\xi'_{M-1}s'_M}
{\tilde R}_{s_{M}^{~}\xi_{M+1}^{~}}^{s'_{M}\xi'_{M+1}} \, ,
\end{equation}
where the improvement for the variational state is performed
via the diagonalization of this $2m^2_{~}$-dimensional matrix.
The eigenvector ${\tilde U}_{\xi_{M-1}^{~}s_M^{~}\xi_{M+1}^{~}}^{~}$ of 
the transfer matrix (Eq.11) is $2m^2_{~}$-dimensional,
and it is possible to rewrite it as~\cite{pwfrg}
\begin{eqnarray}
{\tilde U}_{\xi_{M-1}^{~}s_M^{~}\xi_{M+1}^{~}}^{~} 
&=&
\sum_{\xi_M^{~}}^m
A_{\xi_{M-1}^{~}\xi_M^{~}}^{s_M^{~}} 
\Omega'_{\xi_M^{~}\xi_{M+1}^{~}} \nonumber\\
&=&
\sum_{\xi_M^{~}}^m
\Omega'_{\xi_{M-1}^{~}\xi_M^{~}}
B_{\xi_M^{~}\xi_{M+1}^{~}}^{~~~s_M^{~}} 
\end{eqnarray}
using the Gramm-Schmidt orthogonalization;
$A_{\xi_{M-1}^{~}\xi_M^{~}}^{s_M^{~}}$ and
$B_{\xi_M^{~}\xi_{M+1}^{~}}^{~~~s_M^{~}}$ are
orthogonal matrices that satisfy Eq.3,
and $\Omega'_{\xi_{M-1}^{~}\xi_M^{~}}$ and
$\Omega'_{\xi_{M}^{~}\xi_{M+1}^{~}}$
are $m$-dimensional diagonal matrices.
(Note that in general $\Omega'_{\xi_{M-1}^{~}\xi_M^{~}}$ is not equal to
$\Omega'_{\xi_{M}^{~}\xi_{M+1}^{~}}$.)  
Therefore, to use the block structure ${\bf B} \bullet {\bf B}$ 
in the finite system algorithm is equivalent to 
write down the variational state using the tensor product (Fig.3)
\begin{eqnarray}
U_{s_{1\cdots}^{~}s_N^{~}}^{(M)} = \!\!\!\!\!
\sum_{\xi_2^{~}\cdots\xi_{N-1}^{~}}^{{\rm at~most~}m} \!\!\!\!
&& 
A_{s_1^{~}\xi_2^{~}}^{s_2^{~}} \cdots
A_{\xi_{M-1}^{~}\xi_M^{~}}^{s_M^{~}} 
\Omega'_{\xi_M^{~}\xi_{M+1}^{~}}
\\
&&
B_{\xi_{M+1}^{~}\xi_{M+2}^{~}}^{~~~~~~s_{M+1}^{~}} \cdots
B_{\xi_{N-1}^{~}s_{N}^{~}}^{~~~~~~s_{N-1}^{~}} \, , \nonumber
\end{eqnarray}
and to improve 
$A_{\xi_{M-1}^{~}\xi_M^{~}}^{s_M^{~}}$ and
$B_{\xi_M^{~}\xi_{M+1}^{~}}^{~~~s_M^{~}}$ {\it directly}
via diagonalization of ${\tilde T}_{\xi_{M-1}^{~}s_M^{~}\xi_{M+1}^{~}}^{
\xi'_{M-1}s'_M \xi'_{M+1}}$. All block spins 
$\xi_2^{~}\cdots\xi_{N-1}^{~}$ are at most $m$-state;
this is the non-negligible difference between 
$U_{s_{1\cdots}^{~}s_N^{~}}^{(M)}$ in Eq.13 and
$V_{s_{1\cdots}^{~}s_N^{~}}^{(M)}$ in Eq.2.

Since $\xi_2^{~}\cdots\xi_{N-1}^{~}$
are less than $m$-state in $U_{s_{1\cdots}^{~}s_N^{~}}^{(M)}$, (Eq.13)
we keep all the non-zero eigenvalues of the density matrix
during the numerical calculation of the finite system algorithm;
we don't cut off any states. As we repeat the improvement for 
the tensors $A_{\xi_{i-1}^{~}\xi_i^{~}}^{s_i^{~}}$ and 
$B_{\xi_j^{~}\xi_{j+1}^{~}}^{~~~s_j^{~}}$ for all $i$ and $j$,
the variational state $U_{s_{1\cdots}^{~}s_N^{~}}^{(M)}$
gradually approaches to its fixed point
$U_{s_{1\cdots}^{~}s_N^{~}}^{~}$, {\it which is not dependent on $M$.}
As a result, unlike $\lambda_{~}^{(M)}$ in Eq.5, the expectation value
\begin{eqnarray}
\lambda'_{~} &=& \sum_{\rm all~indices}^{~}
U_{s'_{1\cdots}s'_N} \, 
T^{s'_{1\cdots}s'_N}_{s_{1\cdots}^{~}s_N^{~}} \, 
U_{s_{1\cdots}^{~}s_N^{~}} \\
&=& 
\sum_{\rm all~indices}^{~}
{\tilde U}_{\xi'_{M-1}s'_M\xi'_{M+1}}^{~} \, 
{\tilde T}_{\xi_{M-1}^{~}s_M^{~}\xi_{M+1}^{~}}^{
\xi'_{M-1}s'_M \xi'_{M+1}} \, 
{\tilde U}_{\xi_{M-1}^{~}s_M^{~}\xi_{M+1}^{~}}^{~} \nonumber
\end{eqnarray}
is not dependent on $M$. Figure 4 shows  
${\rm ln} \left(\lambda'_{~}\right)$
of the Ising model; we use the same
system size and parameter as those in Fig.2. 
As we have expected, $\lambda'_{~}$ calculated from Eq.14
does not show any $M$ 
dependence. It should be noted that $\lambda'_{~}$
is always larger than $\lambda_{~}^{(M)}$, which is shown 
for comparison.  Since both $\lambda_{~}^{(M)}$ and
$\lambda'_{~}$ are variational lower bound for the partition function
per row ($=$ unit transfer), it is apparent that $\lambda'_{~}$ is
better than $\lambda_{~}^{(M)}$ in this case. 

We have used the block structure ${\bf B} \bullet  {\bf B}$ 
to obtain $\lambda'_{~}$
after we  calculated $\lambda_{~}^{(M)}$ using the 
standard block structure ${\bf B} \bullet \bullet  {\bf B}$.~\cite{after}
The additional calculation for ${\bf B} \bullet  {\bf B}$ 
is not time consuming at all, because the 
matrix dimension of ${\tilde T}_{\xi_{M-1}^{~}s_M^{~}\xi_{M+1}^{~}}^{
\xi'_{M-1}s'_M \xi'_{M+1}}$ in Eq.11 is smaller than
${\tilde T}_{\xi_{M-1}^{~}s_M^{~}s_{M+1}^{~}\xi_{M+2}^{~}}^{
\xi'_{M-1}s'_M s'_{M+1}\xi'_{M+2}}$ in Eq.6, and  the finite
system process can be done very rapidly with the use of the
wave function renormalization.~\cite{Whacce,pwfrg}
Thus, treating ${\bf B} \bullet  {\bf B}$, one can increase the numerical 
precision in DMRG without increasing computational 
time so much.

One might insist that the difference between 
$V_{s_{1\cdots}^{~}s_N^{~}}^{(M)}$ in Eq.2 and
$U_{s_{1\cdots}^{~}s_N^{~}}^{~}$ in Eq.13 is negligible
if $m$ is sufficiently large. The statement is correct.
We have to keep in mind, however, that occasionally
people are suffering from increasing $m$ in order to keep
numerical precision. It is worth improving
DMRG {\it without} increasing $m$,~\cite{twoclass} because  
to keep large $m$ is difficult for complicated models.

It is straightforward to introduce the ${\bf B} \bullet  {\bf B}$ structure
to the infinite system algorithm in order to recover the 
translational invariance of the variational state
in the thermodynamic limit. The way is, as we have done
for the finite system algorithm, first to employ 
the block structure ${\bf B} \bullet \bullet  {\bf B}$ and to perform
the infinite RG processes till the renormalized wave
function converges,~\cite{ostnew} and then to use use ${\bf B} \bullet 
{\bf B}$ to perform additional infinite  processes.
In this way, the translationally invariant variational
state by \"Ostlund-Rommer~\cite{Ost} is obtained numerically via DMRG.

We thank K.~Okunishi and Y.~Hieida for discussions
about the tensor product formulation, and I.~Peschel for
valuable discussion at the Max-Planck-Institute f\"ur
Physik komplexer System in Dresden. We thank
T.~Tonegawa for allowing us to use his
IBM RISC station type 560 and 595 for this research.

\newpage
\leftline{\large \bf  Figure Captions}

\begin{itemize}
\item[Fig.1] Graphical representation of (a) the variational state, (Eq.2)
 (b) the variational eigenvalue of the transfer matrix, (Eq.5)
and (c) the renormalized transfer matrix. (Eq.6) 
Circles and squares denote bare and renormalized spin variables,
respectively. We use black marks when the variables are
summed up in the corresponding equations.
\item[Fig.2] The logarithm of $\lambda_{~}^{(M)}$, which is the eigenvalue
of the renormalized transfer matrix
${\tilde T}_{\xi_{M-1}^{~}s_M^{~}s_{M+1}^{~}\xi_{M+2}^{~}}^{
\xi'_{M-1}s'_M s'_{M+1}\xi'_{M+2}}$ of the Ising model at the
critical temperature when $N = 200$ and $m = 8$.
Open boundary condition is chosen.
\item[Fig.3]
The graphical representation of the variational state
$U_{s_{1\cdots}^{~}s_N^{~}}^{(M)}$ defined in Eq.13.
The finite system algorithm for ${\bf B} \bullet  {\bf B}$
gradually improves the state, and finally 
$U_{s_{1\cdots}^{~}s_N^{~}}^{(M)}$ lose the $M$ dependence.
\item[Fig.4]
The logarithm of $\lambda'_{~}$, which is the eigenvalue of
${\tilde T}_{\xi_{M-1}^{~}s_M^{~}\xi_{M+1}^{~}}^{
\xi'_{M-1}s'_M \xi'_{M+1}}$ for the Ising model at the
critical temperature when $N = 200$ and $m = 8$.
Unlike ${\rm ln} \left(\lambda_{~}^{(M)}\right)$ shown for comparison,
${\rm ln} \left(\lambda'_{~}\right)$ 
shown by circles does not show any $M$ dependence.
\end{itemize}


\begin{thebibliography}{99}
\bibitem{bnd} The fact does not contradict to the variational principle in
DMRG. The bond energy except at the center of the system is higher than
the true energy per site, thus the variational principle is not violated.
\bibitem{sym} We have assumed that the transfer matrix
     is symmetric. For an asymmetric transfer matrix, we have
     to prepare the dual pair ($\sim$ conjugate) of ${\tilde V}$.
\bibitem{notd} Precisely speaking, $\Omega$ does not have to be
     diagonal. We have shown the simplest case.
\bibitem{kept} We choose such a small $m$ in order to clearly
     visualize the $M$-dependence in $\lambda^{(M)}_{~}$.
\end{thebibliography}
\end{document}